\input harvmac

\vskip 1cm

 \Title{ \vbox{\baselineskip12pt\hbox{ Brown Het-1211 }}}
 {\vbox{
\centerline{ Duality and Combinatorics     }
\centerline{ of Long Strings  in ADS3    } }}   

\centerline{$\quad$ { Mihail Mihailescu  and Sanjaye Ramgoolam}}
\smallskip
\centerline{{\sl  }}
\centerline{{\sl Brown  University}}
\centerline{{\sl Providence, RI 02912 }}
\centerline{{\tt mm,ramgosk@het.brown.edu}}
 \vskip .3in

   The counting of long strings in ADS3, in the context 
   of Type IIB string theory on $ADS_3 \times S^3 \times T^4$,
   is used to exhibit the action of 
  the duality group $O(5,5;Z)$, and in particular its
  Weyl Subgroup $S_5 \bowtie Z_2$, in the non-perturbative
  phenomena associated with continuous spectra of states in these
  backgrounds. The counting functions are related to states in Fock spaces.
  The symmetry groups
  also appear in the structure of  compactifications 
  of instanton moduli spaces on $T^4$. 
 
\noblackbox

\lref\asprev{P. Aspinwall ``String theory and K3 surfaces,'' hepth/9611137. } 
\lref\martlars{  F. Larsen and E. Martinec, `` $U(1)$ charges and moduli 
   in the $(4,0)$ theory,'' JHEP 9911 (1999) 002. } 
\lref\rwal{ S. Ramgoolam, D. Waldram, ``Zero branes on a compact orbifold,''
 JHEP 9807(1998) 009. }
\lref\nik{ Nikulin }
\lref\fot{ M. Fukuma, T. Oota, H. Tanaka, 
            ``Weyl groups in $ADS_3/CFT_2$,'', hepth-9912010. } 
\lref\amikh{ Andrei  Mikhailov, ``D1-D5 system and non-commutative geometry,''
             hepth/9910126. } 
\lref\sw{ N. Seiberg, E. Witten, ``The D1-D5 system and singular CFT,''
              JHEP 9904 (1999) 017 } 
\lref\dijk{ R. Dijkgraaf, ``Instanton Strings and Hyperkahler geometry,''
             Nucl. Phys. B543 (1999) 545-571. }
\lref\mms{  J. Maldacena, J. Michelson, A. Strominger, 
            ``Anti-de-Sitter Fragmentation'' JHEP 9902 (1999) 011. } 
\lref\dtor{  Z. Guralnik and S. Ramgoolam ``Torons and 
           D-brane bound states,'' Nucl. Phys. B499 (1997) 241-252. } 
\lref\hatay{ A. Hashimoto and W. Taylor, ``Fluctuation Spectra of Tilted
                      and Intersecting D-branes from the Born-Infeld 
                      Action,'' Nucl. Phys. B503 (1997) 193-219.  } 
\lref\brz{ M. Berkooz, M. Rozali, N. Seiberg, ``Matrix Description 
 of M-Theory on $T^4$ and $T^5$,'' Phys. Lett. B408 (1997) 105-110 }  
\lref\malda{ J. Maldacena, { \it  The large N limit of superconformal
                             field theories and supergravity,}
                      Adv.Theor.Math.Phys.2: 231-252, 1998,
                       hepth/9711200 }     
\lref\jr{  A. Jevicki, S. Ramgoolam,
   { \it Non commutative gravity from the ADS/CFT
          correspondence,}  JHEP 9904 (1999) 032,  hep-th/9902059 }  
\lref\wad{J.R. David, G.M.Mandal, S. Wadia,  
``Absorption and Hawking Radiation of minimal and fixed scalars, 
 and ADS/CFT correspondence, '' hepth/9808168, 
Nucl. Phys.B544 (1999) 590-611 } 
\lref\jmr{ A. Jevicki, M. Mihailescu and S. Ramgoolam,
``Gravity from CFT on $S^N(X)$: Symmetries and Interactions''
hepth/9907144.  } 
\lref\mih{ M. Mihailescu, ``Correlation functions for chiral primaries
            for $D=6$ supergravity on $ADS_3 \times S^3$,'' hepth/9910111. } 
\lref\wadi{ J.R.David, G. Mandal, S. Wadia, ``D1/D5 moduli in SCFT 
and Gauge theory and Hawking Radiation,'' hepth/9907095.  }
\lref\thft{ G. 't Hooft, ``Some twisted self-dual solutions of the Yang-Mills
equations on a hypertorus,'' Commun. Math. Phys. 81, 1981, 267-275.  }
\lref\malstro{  J. Maldacena, 
A. Strominger,{ \it AdS3 Black Holes and a Stringy
           Exclusion Principle,}  JHEP 9812 (1998) 005, hep-th/9804085 } 
\lref\ahberk{ O. Aharony, M. Berkooz, ``IR dynamics of d=2, N=(4,4) 
 gauge theories and DLCQ of little string theories,'' hepth/9909101. }  
\lref\dk{S.Donaldson and P.Kronheimer, ``The geometry of four-manifolds,'' 
 Oxford Mathematical Monographs, 1997 }  
\lref\fks{  S. Ferrara, R. Kallosh, and A. Strominger, {\it $N=2$
Extremal black holes,} Phys. Rev. {\bf D52} (1995) 5412--5416,
hep-th/9508072. } 
\lref\obpi{ N. Obers, B. Pioline, ``U-duality and M-Theory,''
 hepth/9809039, Phys. Rept. 318, (1999) 113-225 }
\lref\stvaf{ A. Strominger and C. Vafa, ``Microscopic Origin 
 of the Bekenstein-Hawking Entropy,'' Phys. Lett. B 379 (1996) 99-104. } 
\lref\calma{ C. Callan and J. Maldacena, ``D-brane approach to Black Hole 
 Quantum Mechanics,'' Nucl.Phys. B472 (1996) 591-610.  }
\lref\hv{ F. Hacquebord and H. Verlinde,
 ``Duality symmetry on $N=4$ Yang Mills Theory 
 on $T^4$,'' Nucl. Phys. B508 (1997) 609.  } 
\lref\susk{ L. Susskind, ``T-duality in Matrix Theory and 
S-duality in Field Theory,'' hepth/9611164 } 
\lref\grt{O. Ganor, S. Ramgoolam, W. Taylor, 
 ``Branes, Fluxes and Duality in Matrix Theory,''
Nucl. Phys. B492 (1997 ) 191.  } 
\lref\nw{ W. Nahm, K. Wendland, 
``A Hiker's Guide to K3 - Aspects of $N=(4,4)$
 Superconformal Field Theory with central charge $c=6$,''
 hepth/9912067. } 
\lref\vafwit{ C. Vafa and E. Witten, ``A strong coupling test 
of S-duality,'' Nucl.Phys. B431 (1994) 3-77. }
\lref\hm{ J. Harvey and G. Moore, ``On the algebras of BPS states,''
  Commun. Math. Phys. 197 (1998) 489, hepth/9609017.  }

\def\cM{ {\cal{M}} } 

\Date{12/99 }

\newsec{ Introduction } 
 
 We  study counting problems 
 related to the phenomena of long strings
 in ADS3, in the context of 
 type IIB string theory on $ADS_3 \times S^3 \times T^4$, 
 which is one of the examples which enters the 
 Maldacena conjecture \malda. This example is of  special 
 interest because the CFT dual is a tractable 
 2D CFT based on an orbifold $S^N(T^4)$. 
 Various  aspects of the operator algebra and correlation functions 
 have been studied for example in \malstro\jr\wad\jmr\mih.
 An important issue that has to be understood 
 better in order to make the orbifold CFT more useful 
 is the precise map between the moduli space of the 
 CFT and that of the spacetime theory. Significant steps 
 in this direction have been made in \martlars\sw\wadi. 
 Closely related to the issue of moduli is the issue of dualities,
 since the the duality group $O(5,5;Z) $ 
 appears in the  description of the 
 moduli space $ O(5,5;Z)\backslash  O(5,5;R)/O(5) \times O(5)$
 of $R^6 \times T^4$ compactifications 
 of type IIB. The ADS3 background is obtained by choosing 
 a string in six dimensions and going to the near-horizon 
 limit. The string can be a bound state of
 one or more of the following :   D-string,   D5-brane wrapped 
 on the $T^4$,  D3-branes wrapped along the 
 two cycles of the torus, NS one-brane ( elementary string) 
 or NS 5-brane wrapped on $T^4$. The allowed charges
 of the string live are vectors in a lattice $\Gamma^{5,5}$ ( described more 
 explicitly in section 2 ). After choosing a vector in $\Gamma^{(5,5)} $
 describing the background string it is often useful to 
 focus on subgroup $O(4,5;Z)$ of $O(5,5:Z)$, see for example 
 \dijk\amikh. For the kinds of questions we will be asking
 it will be interesting to look at subgroups 
 of $O(5,5:Z)$  which may or may not 
 belong  to $O(4,5;Z)$. 

 A very interesting class of phenomena 
 in ADS3, not directly accessible from 
 the orbifold model at its free point, 
 was studied  in \mms\sw. This involves, in the 
 simplest case, an ADS3 background obtained from 
 the near horizon limit of $Q_1$ D1 branes and 
 $Q_5$ D5-branes. This system allows a D1 ( or D5 )  
 brane to split off and grow to infinity at finite
 cost in energy. A semiclassical calculation  
 (valid for large $Q_1,Q_5$ ) 
 of the split string worldvolume action 
 shows that the string worldvolume theory 
 is a Liouville theory, which is known to have 
 a continuous spectrum above a threshold.
 The threshold is at $Q_5/4$ if we have  a split 
 D1 string, and $Q_1/4$ if we have a split D5-string, 
 The spectrum of the Hamiltonian in ADS3 therefore
 has a continuum starting at $Q/4$ where $Q= min (Q_1,Q_5)$. 
 This formula has the very interesting symmetry 
 under exchange of $Q_1$ and $Q_5$, which is in 
 $O(5,5;Z)$ but not in $O(4,5;Z)$.

 These long string phenomena correspond to
 the splitting   $(Q_1,Q_5) = ( Q_1-1,Q_5) + (1,0)$ 
 or $(Q_1,Q_5) = ( Q_1,Q_5-1) + (0,1)$. The conditions  
 for such splittings $Q= \sum q^{(i)} $ to be BPS were
 described in \sw\  in terms of the geometry of 
 lattices and 5-planes and $R^{(5,5)}$ ( we review this in section 2 ). 
 For generic moduli there are no BPS splittings. 
 An interesting class of splittings happens  when the 
 NS sector B-field moduli and the RR sector 
 C-field moduli are set to zero (  For the bulk  
 of this paper we will work with  the $B=C=0$ condition, with off-diagonal 
 components of the metric set to zero,  and  
 we  get back to non-zero $B,C$ in section 7 ). 
 The same symmetry under exchange of 
 $Q_1$ and $Q_5$ can be seen in the simple exercise of 
 counting of the number of distinct splittings 
 of the $(Q_1,Q_5)$  system. The counting is not symmetric 
 under exchange of $(Q_1,Q_5)$ with $(Q_1Q_5,1)$. 

 In this paper we study generalizations 
 of the counting of BPS splittings, and 
 we find that the Weyl group of $O(5,5; Z)$, 
 which is generated by the symmetric group 
 $S_5$ and a $Z_2$, indicated by writing the Weyl 
 group as $S_5 \bowtie Z_2$, has an interesting 
 action on the counting functions associated 
 with BPS splittings. 
 It is natural to expect 
 that it is also a symmetry of the more detailed
 dynamics of continuous spectra associated with such splittings. 
  Note that we are discussing splittings for systems 
 where the charges have no common factor. As mentioned in 
 \sw\ when the charge vector is non-primitive there are 
 BPS splittings for arbitrary moduli. 
Section 2 is a review of relevant background. 
 Sections 3-5 deal with splittings of
 different kinds of charges. 
 Section 6 discusses the connections of these BPS 
 brane separation problems and the associated symmetries 
 with compactifications of instanton moduli spaces. 
 Section 7 discusses some aspects of splittings 
 beyond $B=C=0$. Section 8 discusses the description 
 from the gauge theory point of view of splittings
 involving NS charges from D-brane systems.

\newsec{ Review }

\subsec{ Backgrounds as 5-planes } 

Choosing a background for IIB theory
on $R^6 \times T^4$ requires choosing 
 a 5-plane in $R^{(5,5)}$, modulo 
discrete identifications \asprev  ( for related discussions 
 see \amikh\dijk\rwal\nw ). 
The 5-plane is 
 spanned by vectors $E^{(0)},E^{(1)},E^{(2)},E^{(3)},E^{(4)} $.
 We use the notation of \amikh\ 
 ( with a minor reshuffling of entries ) 
 for the vectors spanning the positive 
 5-plane $ \Theta $ in $R^{(5,5)}$. 
\eqn\fivp{\eqalign{ 
&  E^{\mu } = (  v^{\mu}; - C.v^{\mu},0  ) 
                       \in R^{4,4} \times R^{(1,1)} \cr 
& E^{4} = ( 0;  \beta, 1 ) \in R^{4,4} \times R^{(1,1)} \cr 
& v^i = (  -B.\omega^{i},0; \omega^{i} ) \in R^{1,1 } \times R^{(3,3)}\cr 
& v^0 = ( \alpha,1;0 ) \in R^{1,1} \times R^{(3,3)} \cr }  } 
Here $\mu$ runs from $0$ to $4$ and $i$ runs 
 from $0$ to $3$. We also have 
 $ \beta = { { 1 \over g_6^2}  - { 1\over 2 }  C.C } $, 
 and $\alpha = V - { 1 \over 2 } B.B $.
 
The charges of strings live in a lattice
$\Gamma^{(5,5)} \in R^{(5,5)}$. 
The charge $Q$ has components 
$(Q_1,Q_5; Q_{12},Q_{34},Q_{13},Q_{42},Q_{14},Q_{23}; N_1,N_5 )$.
Here $(Q_1,Q_5)$ are the charges from D1 strings and wrapped
D5 strings, and $(N_1,N_5)$ are the charges from 
NS strings and wrapped NS 5-branes. 
The $Q_{ij}$ are the charges of strings obtained 
by wrapping D3 branes on the $(ij)$ cycle.   
The bilinear form on $\Gamma^{(5,5)}$ evaluated on two vectors 
$Q^{(1)} = (Q_1^{(1)},Q_5^{(1)}; Q_{12}^{(1)},Q_{34}^{(1)},
Q_{13}^{(1)},Q_{42}^{(1)},Q_{14}^{(1)},Q_{23}^{(1)};N_1^{(1)},N_5^{(1)} )$
and $Q^{(2)} = (Q_1^{(2)},Q_5^{(2)}; Q_{12}^{(2)},Q_{34}^{(2)},
Q_{13}^{(2)},Q_{42}^{(2)},Q_{14}^{(2)},Q_{23}^{(2)};N_1^{(2)},N_5^{(2)} )$, 
 is
\eqn\bilfrm{\eqalign{
&  ( Q^{(1)}, Q^{(2)} ) = Q_1^{(1)}Q_5^{(2)} + Q_1^{(2)}Q_5^{(1)}
           + Q_{12}^{(2)}Q_{34}^{(1)} +Q_{12}^{(1)}Q_{34}^{(2)}  \cr 
&           + Q_{13}^{(1)}Q_{42}^{(2)} +Q_{13}^{(2)}Q_{42}^{(1)}
           + Q_{14}^{(1)}Q_{23}^{(2)} + Q_{14}^{(2)}Q_{23}^{(1)}
           + N_1^{(1)}N_5^{(2)} +  N_1^{(2)}N_5^{(1)} \cr }}

For a rectangular torus ( i.e off-diagonal 
 components of the metric set to zero )
and  with vanishing 
$B$ and $C$ fields, we have 
\eqn\fivps{\eqalign{  E^{0} & = ( V,1;0; 0,0) \cr 
             E^{i} & = ( 0,0; \omega^{i}; 0,0 )\cr 
             E^4 & = (0, 0; 0;  { 1\over g_6^2},1  ) \cr } }  
We can choose : 
\eqn\omeg{\eqalign{ 
  & \omega^{1} = R_1R_2 dx_1\wedge dx_2 + R_3R_4 dx_3 \wedge dx_4 \cr 
   & \omega^{2} = R_1R_3 dx_1 \wedge dx_3 - R_2R_4 dx_2 \wedge dx_4 \cr 
   & \omega^{3} = R_1R_4 dx_1 \wedge dx_4 + R_2R_3 dx_2 \wedge dx_3 \cr }}
$R_i$ are the circumferences of the circles. 
$x_i$ are variables with periodicity $1$. 
 With these formulae the tension  $ T(Q)$ ( in units of $1/g_6$,
 where $g_6$ is 
 the six-dimensional 
 string coupling ),  is given by 
 $ T(Q)= Q_{+}  $, where $Q_+ $ is the projection 
 of $Q$ to the positive 5-plane. With  these expressions, 
 we can recover the mass formulae of \martlars. 

 Since physical quantities are 
 given by projections of 
 vectors in $\Gamma^{(5,5)}$ to 
 the 5-plane, the space of physically inequivalent 
 vacua is obtained by modding out by the symmetries
 of  $\Gamma^{(5,5)}$ :
  $ O(5,5;Z) \backslash O(5,5;R) / O(5,R) \times O(5,R) $. 
 
 When we choose 
 a charge for the string in 6 dimensions 
 living in $\Gamma^{(5,5)}$, having the property 
 $Q^2 > 0$, the attractor equations \fks\ imply that  
 the near horizon moduli satisfy 
 the condition that $Q$ is parallel to the 
 5-plane \dijk\sw.

\subsec{BPS splitting into multiple parts } 

 The condition for a splitting 
   $Q = q^{(1)} + q^{(2)} $ 
   to be BPS can be expressed by saying that 
   the projection of $q^{(1)}$ or $q^{(2)} $ to $\Theta$ 
   is proportional to $Q$ \sw.
   The projection has to be non-negative 
i.e $ q.Q \ge 0$. 

 The density of states in the continuum 
 associated with the splittings into multiple 
 summands will be larger than in the case of two summands. 
  The condition 
  for these splittings to be BPS can be obtained 
  as in \sw.   
  Consider a vector $ Q \subset \Gamma^{(5,5)}$ of charges which 
  can be written as $ Q = q^{(1)} + q^{(2)} + \cdots q^{(l)}$,  
  satisfying the condition that the projections 
  to the positive 5-plane obey
  $ | Q_{+} | = |  q^{(1)}_{+} | + | q^{(2)}_+ | + \cdots |q^{(l)}_+|.  $ 
 The near horizon geometry satisfies $ Q_{+} = Q$.  
 These conditions can be satisfied if each vector 
 $q^{(i)}$ in the decomposition has 
 a projection to the 5-plane,  $q^{(i)}_{+} $  
 which  is parallel to $Q$.

If we start with a system of charges
 $(Q_1,Q_5)$ and study its splittings at 
$B=C=0$ with generic radii, we get    
a  set of splittings sitting  in a 
   $ \Gamma^{1,1}$ lattice. 
   The symmetry of this lattice 
   includes exchanging $(Q_1,Q_5 ) \rightarrow (Q_5,Q_1)$,
   but does not include
   $ ( Q_1,Q_5 ) \rightarrow (Q_1Q_5, 1)$. 
  The duality group $O(\Gamma^{5,5}) $
 does allow us to map the system 
 $(Q_1,Q_5)$ to both $(Q_5,Q_1)$ and to 
 $(Q_1Q_5,1)$.
 The first kind of map allows us to start 
 from $B=C=0$ and end with $B=C=0$.
 The second does not \martlars. 
   There is also S-duality symmetry which is often discussed
   and exploited. It allows us to relate
   the physics of  backgrounds containing NS-NS fields,  
   to  the physics of backgrounds with 
 RR charge. It preserves the $B=C=0$ condition. 
 
 We will study splittings with the condition 
 $B=C=0$, torus rectangular, and identify counting functions
 describing the splittings. These functions  exhibit 
 symmetries in the Weyl group, not too surprisingly since
 this is  known to preserve 
 these conditions on the moduli \obpi. 
 In section three we look at splittings of
  the charge vector  $(Q_1,Q_5; \vec 0; 0,0) $
 system. Section 4 deals with the splittings of 
 $(Q_1,Q_5; Q_{12},Q_{34},0,0,0,0;0,0)$ system. 
 Section 5 describes splittings of 
 $(Q_1,Q_5;Q_{12},Q_{34},Q_{13},Q_{42},Q_{14},Q_{23};N_1,N_5)$.

\subsec{ Liouville theory on long string }

 A Liouville description of long strings was given in \sw.
 Further discussion from a 2D gauge theory point of view
 appeared in \ahberk.  
 We have  a pair of branes, one with charge
 $q$ and the other with charge $Q-q$ splitting from one with charge $Q$. 
 The tensions of the split strings also add up to the tension 
 of the string of charge $Q$. 
A Liouville theory on the long string was derived
\eqn\liouv{
S=T r_0^2 \int \sqrt{g} ((\partial \Phi)^2 + \Phi R)
}
where $T$ is the tension of the brane and $r_0$ is the radius of $AdS_3$.
The Liouville field $\Phi$ is related to the radial direction. After rescaling
the field, the above action is a Liouville action having the background charge 
${\cal Q}=\sqrt{4 \pi T r_0^2}$. The conformal field theory has a 
central charge 
$c_{Liouville}=3 {\cal Q}^2$ and a massgap $\Delta_0={{\cal Q}^2 \over 8}$. 
We assume now that $Q^2 >> q^2$ and that the $AdS$ geometry is given by $Q$. 
Using the same arguments leading to $\liouv$ we obtain using $T(q)=qQ/|Q|$ and
$r_0^2=2 \pi |Q|$ that ${\cal Q}=2qQ$. The previous expression reduces to the 
expected one in the case of $q$ being a $D1$ string, 
i.e giving a central charge of $6Q_5 = 6(Q_1Q_5) - 6(Q_1-1)(Q_5)$.  
\par
In general, we may expect 
 the string of charge $q$ to  see  a geometry given by 
$Q-q$. Interpreting the Liouville theory as living 
on an ``interaction string'' whose central charge 
 is the difference between $3Q^2$ and $3 ( Q-q)^2 + 3 q^2$, we have 
\eqn\centr{
{\cal Q}=2q(Q-q), ~~~c_{Liouville}=6q(Q-q)}
Using the finite $Q_5$ result in \sw\
 we would expect by U-duality that a split 
 string of charge $q$ would lead to  central charges, and thresholds
 given by : 
\eqn\resd{\eqalign
{& {\cal Q}=2q(Q-q),\cr
 & c=3 {\cal Q}^2,\cr
 & \Delta_0= {({\cal Q}-1)^2 \over {4 {\cal Q}}}.\cr }}

\newsec{ Counting of splittings of $(Q_1,Q_5; \vec 0;0,0)$ } 
 We  describe the BPS splittings
 of the system $(Q_1,Q_5; 0;0,0)$.
 We are always working with $B=C=0$ 
 and rectangular tori, unless explicitly 
 specified otherwise. 
 We  first consider  generic radii, and then rational  radii,
 where we further specialize to $R_i = g_6^2 =1$. 

\subsec{ Splittings of $(Q_1,Q_5; \vec 0; 0,0 )$ for generic  radii. } 
 We would like to perform a detailed
 counting of the splittings for a given charge. 
 Consider the charge $(N, 1)$. We can split 
 it as follows : 
\eqn\spl{ (N, 1 ) = ( q, 1 ) + ( q^{\prime}, 0) } 
where $q + q^{\prime } = N$. The  
possible choices of $q$ range from $0$ to  $N -1$, so there 
 are $N$ of them. 
 For small $q^{\prime}$ such BPS splittings 
 lead to a continuum of states as explained in 
 \sw. There are also splittings where we decompose 
 $(N,1)$ into a sum of more than two vectors. 
 These will  also contribute to the Hamiltonian 
 for string theory on $ADS3$ a continuum of states
 ( with the  higher density of states associated 
  with a multistring system as opposed to a two-string system ). 

 To formulate  physically sensible counting rules 
 we have to decide whether a subsystem like 
 $(1,0) + (1,0)$ should be counted as identical to 
 $(2,0)$ or as different. The $(1,0)$ system 
 with unit D1-charge can be mapped to an elementary 
 string. Since we understand the perturbative 
 multi-string Hilbert space, we know that in the sector 
 with winding number two, we can have two singly wound states
 or a doubly wound state. So $(1,0)+ (1,0)$ should be counted differently. 
 Similarly we can argue that a string with charges 
 $(2,2)$ should be counted
 as distinct from as string $(1,1) + (1,1)$. 
 By duality this is related to the fact that with 
 winding number $2$ and momentum $2$, we can have 
 a single string with these quantum numbers 
 or two separate strings with these quantum numbers.

   We have two kinds of splittings of 
   the charge vector $(N,1)$. The first has the 
 form   $( N , 0 ) = ( 0, 1) + (N_1, 0 ) + \cdots ( N_l, 0) $, 
 where $N$ is being partitioned into $l$ non-zero parts
 $N = N_1 + N_2 + \cdots N_l $. 
  The second kind of splitting takes the form 
  $ ( N , 0 ) = ( N_1, 1) + (N_2, 0 ) + \cdots ( N_l, 0) $.
  For each partition of $N$ we can place the $1$ 
 in a number of different ways, but if the partition 
 contains some integer more than once, placing the 
 $1$ next to different copies of the integer 
 should not be counted more than once. 
 As we see in the next paragraph this is automatically 
 taken into account by a Fock space description in terms
 of bosonic oscillators.   
 Equivalently for each  partition of 
 $N$, there are $k$ different ways of placing 
 the $1$ where $k$ is the number of distinct integers 
 in the partition.  
 So the total number of splittings of the second kind  is equal 
 to the sum of $k$ over all the partitions of $N$.

 This counting can be written in terms of 
 a function 
\eqn\genfunc{ 
\tilde P(x,t) = \prod_{l=1}^{\infty} { 1 \over ( 1-x^l) } { 1 \over 
   ( 1- t x^l) } } 
Let $ P(x,t) = \sum_{N,m=0}^{\infty} P (N,m) x^N t^m $. 
 The number of splittings is
\eqn\numspl{ 
 N_s (N,1) = P(N,0) + P(N,1) } 
Note that $P(N,0) $ is just the number 
 of partitions of $N$, and it counts 
 the number of splittings of the 
 form $(N,1) = (0, 1) + ( N_1, 0) + (N_2, 0) + \cdots (N_k, 0) $. 
 The second term $P(N,1)$ counts the 
 number of splittings of the form 
   $  (N,1  ) = ( N_1, 1) + (N_2, 0) + \cdots (N_k, 0)$. 
This counting can be described in Fock space 
 language. Consider the oscillators $ \alpha_{-k,l} $
 where $k$ can take values from $1$ to $N$, 
 and $l $ is a discrete index which can take values 
 $0$ or $1$. $P(N,0)$ counts the number of states
 in a Fock space where all the $l$ take the value 
 $0$, and the oscillator indices $k$ add up 
 to $N$.
\eqn\fksp{ \alpha_{-N_1,0} \cdots \alpha_{-N_k,0 } |0> } 
  $P(N,1)$ counts the number of states 
 where the $l $ values add up to $1$, which is the 
 same as saying that one of the $l $ values 
 takes the value $1$. One of these states is : 
\eqn\fkspi{ \alpha_{-N_1,1} \cdots \alpha_{-N_k,0} |0> } 
 The bosonic statistics of the $\alpha$ oscillators
 guarantees that associating the $1$ 
 with different copies of the same $ N_i$ overcounted. 
  A more economical way of writing the number 
  of splittings is to use the generating function : 
\eqn\genfunc{ 
  P_1( x_1,x_2 ) = \prod_{k,l } { 1 \over ( 1 - x_1^{k} x_2^{l} )} } 
where $k $ and $l$ extend from $0$ to $\infty$, 
 except that $k =l =0$ is not allowed. 
Define the coefficient $P_1(n,m)$ as the coefficients 
 appearing in the expansion :
\eqn\coef{ P_1(x_1,x_2) = \sum_{n,m} P_1(n,m) x_1^n x_2^m } 
The number of splittings is just $P_1(N,1)$.
 It is clear that $P_1(N,1) $ is equal to $P_1(1,N)$.

 We can also count the number of ways 
 of splitting a more general charge $(Q_1,Q_5)$. 
 We can write the number of distinct 
 splittings  $N_s(Q_1,Q_5)$ in terms of the coefficients
 of this quantity as 
 follows: 
\eqn\coefqt{ N_s (Q_1,Q_5) = P_1 (Q_1,Q_5) } 
 This number is equal to the 
number of states in a Fock space 
of the form 
\eqn\fone{ \alpha_{-(k_1,l_1)}  \alpha_{-(k_2,l_2)}
 \cdots \alpha_{-( k_m,l_m)}  |0> } 
where the integers $k_1$ through $k_m$ add up to 
  $Q_1$ and the integers $l_1$ through 
 $l_m$  add up to $Q_5$. Oscillators  
 of the form $  \alpha_{-(k,0) }$ or $\alpha_{-(0,l)}$  are allowed 
 but  $  \alpha_{-(0,0) }$ is not allowed. 
It is clear that we have the symmetry : 
\eqn\sym{ N_s ( Q_1,Q_5) = N_s ( Q_5,Q_1).  }

To make the counting even more 
explicit, we can restrict to splittings
 into two parts only, to get 
for the system $(Q_1Q_5,1)$ a number
$Q_1Q_5$. For the system $(Q_1,Q_5)$ we get
${( Q_1+ 1)( Q_5+ 1 ) \over 2} - 1 $.
These formulae clearly show that there is a symmetry 
under exchange of $Q_1$ and $Q_5$, but 
 no symmetry under exchange 
of $(Q_1,Q_5)$ into $(Q_1Q_5,1)$. 

Note that in all of the above counting we have not included
 splittings of the form  
 $ (Q_1,Q_5) = ( -q_1^{\prime} , -q_5^{\prime}) +  
    (q_1 +q_1^{\prime } , q_5+q_1^{\prime} )$, with 
 $q_1^{\prime}, q_5^{\prime} $ positive. 
 From the connection to instanton moduli 
 spaces discussed in a later section, this is
 a natural restriction to consider. 
 For a splitting of this form, when $q_1^{\prime},q_{5}^{\prime}$
 are positive and small,  
 the arguments of \sw\ 
 show that the energy required to create such a long  string 
 is infinite and the system is not BPS. 
 A duality invariant way of stating this 
 restriction is that we are considering 
 splittings $ Q = q^{(1)} + \cdots q^{(l)} $ 
 where the projections of $q^{(i)}$ to the 5-plane 
 are not only proportional  to $Q$ but that the  
  constant of proportionality is positive
 i.e $q^{(i)}.Q \ge 0$.

\subsec{ Refinements of counting }
 
Note that to keep the formulae
simple, we have counted splittings by merely 
counting the number of distinct string charges
that can appear in the splittings of a given charge.  
One may also weight  the splittings by the  number of distinct 
states associated with the ground states of the 
split strings. If we choose to do such a counting, 
we should have, for a charge $(k,l)$,  a 
 multiplicity of $p_{16} ( kl ) $, where $p_{16} ( kl )$
 is the number of states at level 
$kl$  in a Fock space with $16$ 
 distinct oscillators. Such a counting can be done by modifying 
 the above partition functions as follows :  
 We replace in \genfunc\ the expression 
${ 1 \over ( 1 - x_1^{k} x_2^{l} )}$ by 
${ 1 \over {( 1 - x_1^{k} x_2^{l} )^{p_{16}(kl)}} }$.
A further refinement would  involve defining 
a partition function which sums   the lowest energy states 
coming from each splitting by weighting with an 
energy dependent exponential.
 This requires further dynamical information on the energy 
 associated with each splitting.  
The simplest counting problem we have considered, 
 which we extend to systems with more charges
 in the subsequent sections,  
suffices for the purpose at hand which is to 
exhibit the appropriate subgroups of the duality group.

\subsec{ Splittings of $(Q_1,Q_5;\vec 0 ;0,0)$  at special radii. }

 Consider the splitting
 of a 
 \eqn\splxt{\eqalign{   (Q_1, Q_5; 0,0, 0,0,  0, 0; 0,0) &\rightarrow 
 q + (Q-q);  \cr  
 q & = (q_1 , q_5 , q_{12} , q_{34},0,0,0,0 ; 0,0) \cr 
 }}
 According to \sw, this is a BPS splitting
 if the projection of $q$ 
 to the 5-plane is proportional to 
 $(Q_1,Q_5)$. Vanishing of 
 the projection along $E^{1}$ 
   leads to the  condition 
\eqn\fcon{ 
{ - q_{12} \over q_{34 } } = { R_1R_2 \over R_3R_4}  }
 This means that such splitting can only 
 happen when   ${ R_1R_2 \over R_3R_4}$ is a rational 
 number.  
 While the small instanton singularity discussed
 in \sw\ requires tuning $B$-fields, 
 these singularities require tuning geometrical 
 moduli. The fact that we need to tune the 
 B-field to zero could be understood from properties 
 of instantons because the instantons are no longer 
 point-like for non-zero generic B-fields.

 It is interesting that these conditions 
 on the geometry can be understood
 from properties of instanton moduli spaces
 ( anti-self dual connections in our conventions). 
 The splitting should correspond to 
 instanton solutions where 
 the connection takes block diagonal form : 
\eqn\blkfrm{ 
              A = \pmatrix {  & . \quad . & . \quad . \cr  
                              &  ~  A^{(1)}~    & . \quad . \cr 
                              & . \quad .& . \quad . \cr 
                              & . \quad .& A^{(2)}   \cr
                              & . \quad . & .  \quad . \cr }}
Here $A^{(1)} $ is a connection 
in $U(q_5)$ with instanton number $q_1$, 
and fluxes
 $$\int { 1 \over 2 \pi } tr F_{12}=  q_{34}, 
    \int  { 1 \over 2 \pi } tr F_{34}=  q_{12} $$
$A^{(2)}$ is a $U(Q_5 - q_5)$ connection  with instanton number
 $Q_1- q_1$, 
and fluxes 
$-q_{12},-q_{34}$.
Consider first 
 the case of  special cases where $q_1q_5 + q_{12}q_{34} = 0 $
( where $q_1$ is the number of anti-instantons, proportional 
 to $ -\int tr F \wedge F $ ).  
 We can realize this as a $U(q_5) \subset U(Q_5) $ 
 instanton configuration 
 built from  constant field strengths $F_{12},F_{34}$. 
 The flux quantization conditions combined
 anti-self duality,
  lead to the rationality conditions on the radii : 
 \eqn\combcons{\eqalign{&   F_{12}R_1R_2 = q_{34} \cr 
              &   F_{34} R_3R_4 =q_{12} \cr 
              &   F_{12} = -F_{34} \cr 
              &  \Rightarrow { R_1R_2 \over R_3 R_4 } 
                   = - { q_{34} \over q_{12} } \cr }} 
 The anti-self duality is required for
 supersymmetry of  $SU(Q_5)$ configurations. 
 The diagonal constant $U(Q_5)$ part is not constrained.
 This follows from the SUSY variation of the world-volume 
 fermions \hm : 
\eqn\susyvar{ \delta_{ \xi, \tilde \xi} \lambda  = 
\xi \Gamma_{\mu \nu} F^{\mu \nu } + \tilde \xi }   
 The diagonal part of the $U(q_5)$ 
  configuration  is part of the  $SU(Q_5)$. 
Such constant field strength solutions (torons ) 
 were written down in \thft\ and their
 connection to D-brane bound states studied 
 in \dtor\hatay. 
 Torons are in this sense closely related 
 to splittings involving $  q^2 =0 $. 
 To get configurations with $q^2 > 0 $, 
 we have to turn on anti-self dual configurations 
 inside the $SU(q_5)$ which can be either 
 point-like or smooth.

 More generally we can have 
 \eqn\splti{
 (Q_1,Q_5;  0,0,0,0,0,0; 0,0) \rightarrow 
  ( q_1 ,q_5;  q_{12}  , q_{34}, q_{13},
  q_{42}, q_{14}, q_{23} ; n_1, n_5 ) + \cdots  } 
 If all the radii are adjusted to be equal to each other, 
 and the six-dimensional coupling also adjusted to $1$ : 
 we now have four  independent parameters
 which can be arbitrary integers, $( q_{12}, q_{13}, q_{14}, n_1) $. 
 The remaining parameters are determined by 
\eqn\conds{\eqalign{ 
& q_{34} = - q_{12} \cr 
& q_{42} = - q_{13} \cr 
& q_{23} = - q_{14} \cr
& n_5 = - n_1 \cr }}
 The norm of the first vector is equal 
 to $  2 (   q_{1} q_{5} - q_{12}^2 - q_{13}^2 
       - q_{14}^2 - n_1^2 )    $.
   The BPS condition requires that this 
   is greater or equal to zero. 
When we realize these configurations 
 in gauge theory, we have a condition 
 $q_5  > 0 $. The condition 
 $q^2 \ge 0$ means that $ q_1 \ge 0 $. 
 Since the $q_1 $ sum to $Q_1$ 
 and the $q_5$ sum to $Q_5$, 
 we have $  0 \le q_1 \le Q_1 $
 and $ 0 \le q_5 \le Q_5 $. The condition 
$q^2 \ge 0$ then only allows a finite number
 of solutions. 

The splittings of this form 
 can be counted using generating  
functions : 
\eqn\newP{ 
 P_2 (x_1,x_2,x_3,x_4,x_5,x_6 ) = \prod_{k,l,m,n,p,r}
 {  1 \over { (  1 - x_1^{k} x_2^{l} x_3^{m} x_4^{n} x_5^{p}x_6^{r} )  }  }}
The product is constrained 
 by 
\eqn\const{\eqalign{
&    k \ge 0 , l\ge 0 \cr 
&   kl - m^2 - n^2 - p^2 - r^2 \ge 0 \cr }}
Defining $P(n_1,n_2,n_3,n_4,n_5,n_6)$ as the coefficient 
of $ x_1^{n_1} x_2^{n_2} x_3^{n_3} x_4^{n_4} x_5^{n_5} x_6^{n_6 } $ in 
$P(x_1,x_2,x_3,x_4,x_5,x_6)$, the desired 
 counting function is 
$P(Q_1,Q_5,0,0,0,0)$.

 At the special radii and $g_6$ we  
 have a large class of splittings
 involving strings with 
 extra 3-anti3 charges and NS1-5 charges
 $(q_1,q_5;  q_{12}, q_{34}, q_{13}, q_{42}, q_{14},q_{23}; n_1,n_5 )$.  
 We saw  that the charges $q_{34},q_{42},q_{23},n_5$ 
 are determined by the charges $q_{12},q_{13},q_{14},n_1$ respectively. 
 There is an $S_4 \bowtie Z_2$ subgroup 
 of the Weyl group which acts on the charges of these split strings.
 The $S_4$ just permutes the four independent charges 
 and the $Z_2$ acts by a reflection $q_{12} \rightarrow -q_{12}$.  
 The counting function $P(Q_1,Q_5)$ of course has the symmetry 
 of exchanging  $Q_1$ and $Q_5$. 
 A $Z_2 \times ( S_4 \bowtie Z_2 )$  subgroup 
 of $( S_5 \bowtie Z_2)$ is therefore manifest here. 
 The first factor is a symmetry acting on the charge
 of the string we start with i.e exchanging $Q_1,Q_5$. 
 The second is a symmetry acting on the strings that can appear 
 in the splitting.  
 By choosing $R_i =g_6^2=1$ we made sure that 
 the  $S_4 \bowtie Z_2$ acts as a symmetry of the Hamiltonian
 for the corresponding ADS background. For more general radii 
 satisfying the rationality conditions, 
 a split string for one gemoetrical 
 modulus is mapped to a split string at another geometrical 
 modulus.

\newsec{ Splittings of a system $(Q_1,Q_5;Q_{12},Q_{34},\vec 0;0,0 )$ }

 Consider first the splittings of  
\eqn\spltng{ 
 (Q_1,Q_5, Q_{12} , Q_{34} ) = (q_1,q_5;  q_{12} , q_{34},0,0,0,0; 0,0 ) 
   + \cdots } 
We again consider rectangular tori 
 with $ B =C=0$. 
 The condition on the  
 near horizon geometry takes the form : 
\eqn\nhgoem{\eqalign{  
&  { Q_1 \over Q_5 } = V  \cr 
&   { Q_{12} R_1R_2 \over Q_{34} R_3R_4 }  =  1 \cr }}
The condition that the projection of the vector 
 $q$ to the five-plane is parallel to $Q$ 
takes the form :  
\eqn\part{ 
 { ( q_{1} Q_{5} + q_5 Q_1 ) \over {Q_1Q_5} } 
=  { ( q_{12} Q_{34} + q_{34} Q_{12} ) \over ( Q_{12} Q_{34} )} }  
If we consider splittings of the 
more general form : 
\eqn\gspl{ 
(Q_1,Q_5, Q_{12} , Q_{34} ) = (q_1,q_5, q_{12} , q_{34}, 
                          q_{13}, q_{42}, q_{14}, q_{23}, n_1,n_5 ) + ... }
we still have the above conditions 
 and we have restritions on the moduli 
 of the form : 
\eqn\resm{\eqalign{ 
& {  q_{13} R_1R_3 \over q_{42} R_4R_2}  = -1\cr 
& { q_{14} R_1R_4 \over q_{23} R_3R_2} = -1 \cr 
& { { n_1} g_6^2  \over n_5 } = -1 \cr }}

Let us first consider splittings of the first kind.
These can be counted as the coefficient of 
$x_{1}^{Q_1} x_{2}^{Q_5} x_{3}^{Q_{12}} x_{4}^{Q_{34}}$  in the 
series
\eqn\secf{ P(x_1,x_2,x_3,x_4) = 
\prod_{q_1,q_5,q_{12},q_{34}} 
{ 1 \over  ( 1 - x_1^{q_1}x_2^{q_5}x_3^{q_{12}} x_{4}^{q_{34} } )} } 
Here the $q$'s obey the conditions in \part, 
 in addition to the usual $q^2 \ge 0$ and $ q.Q \ge 0$. 
A class of solutions of these conditions
( we haven't proved that they give a complete set 
 of solutions ) can be written as
\eqn\setso{  q = \lambda_1 q_{(1)} + \lambda_2 q_{(2)} } 
where $ q_{(1)} = 
( { Q_1 \over \lambda_L} , 0; 
 { Q_{12} \over \lambda_L},0,0,0,0,0;0,0)$
and 
$q_{(2)} =
 (  0,{ Q_5 \over \lambda_R} , 0; 0,{ Q_{34} \over \lambda_R},0,0,0,0,0;0,0)$.
Here $$\lambda_L = \gcd (Q_1, Q_{12}) ; \lambda_{R} = \gcd (Q_5,Q_{34})$$
and $$0 \le \lambda_1 \le \lambda_L; 0 \le \lambda_2 \le \lambda_R. $$ 
Splittings coming from this class of solutions 
 can be counted as the coefficient of 
$x_1^{\lambda_L} x_2^{\lambda_R}$ 
in the series : 
\eqn\sers{ \prod_{\lambda_1,\lambda_2} 
{ 1\over (1-x_1^{\lambda_1} x_2^{\lambda_2} ) }  } 
This counting function 
$N_s(Q_1,Q_5,Q_{12},Q_{34})$ is symmetric 
 under a group $S_2 \bowtie Z_2$. The $S_2$ exchanges 
 the $(Q_1,Q_5) $ pair with the $(Q_{12},Q_{34}) $
 pair. The $Z_2$ generator can be taken to be the
exchange of $(Q_1,Q_5) $ to $(Q_5,Q_1)$. 

As in the previous section 
we can proceed to the case where the other 
ratios of radii and the coupling  are adjusted according to 
\resm\ and we can further specialize to the 
case where all $R_i = 1 =g_6^2$, to find 
a group  $S_3 \bowtie  Z_2$ acting on  the allowed extra charges.

\newsec{ Splittings of a system 
$(Q_1,Q_5,Q_{12},Q_{34},Q_{13},Q_{42},Q_{14},Q_{23},N_1,N_5)$ } 

 The condition that this charge be parallel 
 to the 5-plane gives : 
\eqn\parcon{\eqalign{ & { Q_1\over Q_5V } = 1\cr 
                      & { Q_{12}R_1R_2 \over Q_{34}R_3R_4 } =1 \cr 
                      & { Q_{13} R_1R_3 \over Q_{42} R_2 R_4} = 1 \cr 
                      & { Q_{14}R_1R_4 \over Q_{23} R_2R_3} = 1 \cr 
                      & {N_1 g_6^2 \over N_5 } = 1 \cr }}
 Consider splittings of the form 
 $ Q \rightarrow q + \cdots $, 
 where the charge $q$ has components 
 $( q_1,q_5; q_{12},q_{34},q_{13},q_{42}, q_{14},q_{23}; n_1,n_5)$.
 We have conditions $q^2 \ge 0$, 
 and requiring that the projection 
 of $q$ to the 5-plane be parallel to 
 $Q$ leads to conditions : 
\eqn\condd{\eqalign{
&    { ( q_1Q_5 + q_5Q_1 )\over Q_1Q_5 } 
 = { ( q_{12} Q_{34} + q_{34}Q_{12} ) \over Q_{12}Q_{34}}  
 = { ( q_{13} Q_{42} + q_{42} Q_{13} ) \over Q_{13} Q_{42} } \cr  
&= { ( q_{14}Q_{23}+q_{23}Q_{14} ) \over Q_{14}Q_{23} } 
 = { (n_1N_5 + n_5N_1 ) \over N_1N_5 }  \cr  }} 
A class of solutions to these equations takes
the form 
\eqn\clssol{\eqalign{ &   \lambda_{1} q^{(1)} + \lambda_2 q^{(2)}, \cr 
&          0  \le \lambda_1 \le \lambda_L \cr 
&          0 \le \lambda_2 \le \lambda_R \cr } }   
where 
$\lambda_L = \gcd ( Q_1, Q_{12}, Q_{13}, Q_{14}, N_{1} ) $ and 
$\lambda_R = \gcd ( Q_5, Q_{34},  Q_{42}, Q_{23}, N_5 )$, 
 and $q^{(1)} = { 1 \over \lambda_{L}}  
(Q_1,0,Q_{12},0,Q_{13},0,Q_{14},0,N_1,0 )$ 
 and $ q^{(2)} = { 1 \over \lambda_{R} }
  ( 0,Q_5,0,Q_{34},0,Q_{42},0,Q_{23},0,N_5  ) $. 
The number of these  splittings 
$N_s(Q_1,Q_5,Q_{12},Q_{34},Q_{13},Q_{42},Q_{14},Q_{23};N_1,N_5) $
 can be counted 
using  the coefficient of $x_1^{\lambda_L}x_2^{\lambda_R} $
in the following series: 
\eqn\fser{ 
 \prod_{\lambda_1,\lambda_2} 
{ 1\over (1- x_1^{\lambda_1} x_2^{\lambda_2} )},  } 
where $\lambda_1 \ge 0 , \lambda_2 \ge 0$, but
$ (\lambda_1,\lambda_2) = (0,0)$ is excluded. 

 The function  
$N_s(Q_1,Q_5,Q_{12},Q_{34},Q_{13},Q_{42},Q_{14},Q_{23};N_1,N_5)$
is invariant under $S_5 \bowtie Z_2$. 
The  $Z_2$ can be taken to be 
the exchange of $(Q_1,Q_5)$ to $(Q_5,Q_1)$. 
The $S_5$ permutes the five blocks : 
 $(Q_1,Q_5)$, $(Q_{12},Q_{34})$, 
 $(Q_{13},Q_{42})$,  $(Q_{13},Q_{42})$ ,  $(Q_{14},Q_{23})$ and  $(N_1,N_5)$.

\newsec{  Beyond $B=C=0$. }

 The  duality group $O(5,5:Z)$ can be used 
 to map the charges $(Q_1, Q_5, 0,0) $ to the 
 charge  $ (Q_1Q_5, 1, 0, 0 ) $.
 An $O(2,2)$ subgroup of the $O(5,5)$ 
 will suffice to do that, as pointed out 
 by \martlars : 
 \eqn\ott{  \pmatrix { & adQ_5^2 & &-bc Q_1^2& &acQ_1Q_5& &bd Q_1Q_5& \cr
                      & -bc &  &ad &  & -ac&  & -bd&  \cr 
                      & ab Q_5& &-abQ_1& &a^2Q_5&  &b^2Q_1& \cr
                       &cd Q_5 &  & - c d Q_1& & c^2 Q_1& & d^2 Q_5& \cr } }
 Here $a,b,c,d$ have been chosen to satisfy 
 the condition $adQ_5 - bcQ_1 =1$.  
 In fact there exists a choice of 
 $c=d=1$  which leads to 
 \eqn\otti{ \pmatrix { & aQ_5^2 & &-b Q_1^2& &aQ_1Q_5& &b Q_1Q_5& \cr
                      & -b&  &a &  & -a&  & -b&  \cr 
                      & ab Q_5& &-abQ_1& &a^2Q_5&  &b^2Q_1& \cr
                       & Q_5 & & -  Q_1& &  Q_1& &  Q_5& \cr } }. 
These matrices allow us to see explicitly 
that these transformations which violate the 
$B=C=0$ condition.

 We can use this map start with the splittings of 
 a $(Q_1,Q_5) $ system to get splittings of 
 a $(Q_1Q_5,1)$ system.  
   The splittings of $(Q_1,Q_5) = (4,3)$
   are studied here. This can be mapped  
   by $O(5,5;Z)$ to the charge $(12,1)$. 
  A first class of splittings of $(4,3)$ involving 
  no extra charges is : 
\eqn\splfst{\eqalign{ (4,3) & \rightarrow (4,0 ) + ( 0,3) \cr 
                            & \rightarrow (4,1 ) + ( 0,2) \cr 
                           & \rightarrow (4,2 ) + ( 0,1) \cr  
                           & \rightarrow (3,0) + (1,3) \cr 
                           & \rightarrow (3,1) + (1,2) \cr  
                           & \rightarrow (3,2) + ( 1,1) \cr 
                           & \rightarrow (3,0 ) + (1,0) \cr 
                           & \rightarrow (2,0) + (2,3) \cr 
                           & \rightarrow (2,1 ) + ( 2,2 ) \cr }}
We also have $12$ splittings of the charge 
 $(12,1)$ into different charges of the 
 form $ ( k,1) + (12-k,0)$ where $k$ ranges from
 $11$  to $0$. When these are mapped to splittings 
 of the $(4,3)$ system using $O(5,5;Z)$ ( in fact 
 $O(2,2;Z)$ suffices ), we get a bunch of splittings 
 of 
\eqn\spltscd{ (4,3) \rightarrow
 (16-k,  k-9 , k-12,k-12 ) + (k-12,12-k,12-k,12-k) } 
 This shows that the class of counting functions 
 etc. that we have described can be 
 used to obtain information about splittings 
 at moduli which go beyond the $B=C=0$ condition. 
  In general the $B,C$ values have explicit 
 dependence on $Q_1$ and $Q_5$, and 
such dependence will show up in the 
 coefficients of marginal operators
 necessary to go from 
the free point of the orbifold description
in terms of $S^{Q_1Q_5 } (T)$ to the point which has a simple 
 description in terms of $(Q_1,Q_5)$ system \martlars. 
  
The data associated with splittings and thresholds
 etc. seems  best viewed as 
living on the space 
\eqn\smla{ \Gamma^{5,5} \times O(5,5;R)/O(5) \times O(5),  } 
 with a manifest $O(5,5;Z)$ symmetry 
 which acts on both charges and moduli. 
 It will be interesting to see if the class 
 of splittings at $B=C=0$, a large class of which 
 we have considered in this paper, exhausts
 all the physics of long strings that one encounters
 as one moves in the space \smla.

\newsec{ Instantons } 

 Properties of instanton moduli 
 spaces have come up in the discussions of the 
 splittings above. Some aspects are developed in more 
 detail here, and some puzzling features discussed. 

 First consider the splittings 
 of $(Q_1,Q_5)$
 which do not involve extra 
 3-brane charges. They correspond to  
 a decomposition of the compactified moduli 
 space of instantons. 
 We will discuss some features \foot{ We thank George Daskalopoulos
 for instructive discussions  on this subject } 
 of the Uhlenbeck compactification \dk. 

 The compactified moduli space of instantons on $T^4$ 
  for gauge group $U(r)$ and instanton 
  number $k$, $\bar \cM_{(r,k)} $, 
  admits a double stratification, labelled by 
  two integers $(f,p)$ which count the number 
  of $ U(1) $ flat connections and the number of 
  point-like instantons respectively. 
 T-duality inverts the volume, exchanges 
 rank and instanton number, and flips the integers  
 $f$ and $p$. 
 \eqn\ds{   \bar \cM_{r,k } = 
 \bigcup_{(f,p)} (  \cM_{r-f,k-p}^{(0)} \times S^p(T) \times S^f(T^* ) )  
 \quad \bigcup ( S^k(T) \times S^r(T^*) )  }  
 The Moduli spaces $\cM^{(0)} $ appearing in the above 
 involve no flat connections and no point-like instantons. 
 The integer $f$ extends from $0$ to $r-2$
 and $p$ extends from $0$ to $k-2$. 
 Naively one may have allowed them to extend
 to $r-1$ or $k-1$, but $\cM_{1,l}$ 
 and  $\cM^{(0)}_{l,1}$  are empty for any $l$.
 In fact such strata would be expected 
 if we want a compactification which knows
 about all the possible split strings, including 
 those of charge $(l,1)$ as explained below.

  Let us consider first the case where $r,k$ are large, 
 and $f,p$ are small. The symmetric product  
 $S^p(T)$ describes $p$ long strings moving 
 on $T$. The sigma model on this stratum contains sectors
 parametrized by partitions of $p$ which describe 
 different ways of partitioning the $p$ long  D1-strings into bound states. 
 Similarly the symmetric product $S^f(T)$ contains sectors
 describing $f$ long D5-strings. 
 A stratum parametrized by $(p,f)$   
 contains the physics of 
 splittings of the type 
 $(Q_1,Q_5 ) \rightarrow ( Q_1-f, Q_5-p) + ( f_1,0) + \cdots (f_l,0) 
   + ( 0, p_1 ) + \cdots (0,p_k) $
 where $p= p_1 + p_2 + \cdots p_k $ 
 and $f = f_1 + f_2 + \cdots f_l $ are partitions 
 of $p$ and $f$. Geometrical structures corresponding to 
$(f,p)$ bound states with both $f$ and $p$ non-zero 
do not seem to appear. Another puzzle appears when we allow
 $p$ to be comparable to $Q_5$. 
 When $p$ is equal to $Q_5-1$ we do not 
 have the corresponding stratum, whereas the algebraic calculation 
 of allowed BPS aplittings 
 continues to allow such splittings.  
 Perhaps other compactifications, e.g the Gieseker 
 compactification,  
 do include the extra strata which would give a more detailed
 correspondence between strata and splittings.

 As discussed in section 3.2, when we consider 
 splittings involving extra 3-brane charges
 we need to consider reducible connections, i.e
 of block diagonal form where each block contains 
 non-zero flux, but the fluxes add up to zero. 
 When the original system contains extra 3-brane 
 charges, we need to start with a bundle 
 of rank $Q_5$ instanton number $Q_1$ and  
 magnetic fluxes $\epsilon^{ijkl} Q_{kl} $. 
 The BPS splitting conditions in \condd\ 
 for example do not allow a point-like instanton
 of charges $(1,0,0,0...)$. This means that for such a choice of 
 moduli and charges, the instanton moduli spaces 
will have no strata corresponding to 
 shrinking instantons. Rather the strata will correspond to 
 solutions of \condd.

\newsec{ Splitting NS charges from D-brane systems }

  The splittings of the 
   $(Q_1,Q_5)$ system which 
   involve extra charges like three branes 
  are easily described in the gauge theory 
 of $U(Q_5)$ which contains magnetic fluxes 
 allowing the description of the appropriate 
 splittings. 
 But the splittings which involve 
 extra charges NS1-NS5 seem to be harder to describe. 
 The NS1-charges are of course described
 by electric fluxes. We don't know how to describe the 
 NS5-charges as a flux in the D-brane gauge theory. 
 This is the famous transverse 5-brane problem 
 in attempts to use Yang Mills as  Matrix theory for 
 compactifications on tori of dimension larger
 than $4$. The transverse 5-brane problem is solved
 by appealing to little string theory \brz\
 but 
 available  descriptions of this theory 
 do not allow an explicit description of these splittings 
 which generalize in a simple way the description  in terms 
 of point-like instantons, flat connections, 
 block-diagonal connections that we have described.
 A description which stays within the 
 confines of moduli spaces of six-dimensional 
 gauge theories, can be given at the expense of giving up 
 the restriction to a fixed rank gauge group. The following 
 remarks have some analogies  to developments  on duality
  in Matrix theory appearing in \susk\grt\hv. 

  An example of a splitting 
   which involves the NS1-NS5 charges is 
 \eqn\spl{  ( Q_1,Q_5;\vec 0 ; 0,0 ) \rightarrow ( q_1, q_5 ; \vec 0 ; 1,-1 ) 
           + (  Q_1-q_1, Q_5-q_5; \vec 0 ; -1,1) } 
 For such a  splitting to exist at $B=C=0$,  we need to adjust the 
 six-dimensional coupling   
 $g_6^2 = 1 $. One way to give a gauge theoretic 
 description of such a system is to map the 
charge $(q_1,q_5; \vec 0 ; 1,-1 ) $
 to $( \tilde q_1,\tilde q_5; \vec 0 ; 0,0)$, 
with $  \tilde q_1\tilde q_5 = q_1q_5 - 1$,   at the cost 
 of turning on some non-trivial RR and/or B-fields. 
 But this is a system which we can describe 
 using $U(\tilde q_5)$ gauge theory with instanton 
 number $\tilde q_1$,  with non-trivial couplings 
 turned on due to the background fields. 
 Similarly we can map 
$  (  Q_1-q_1, Q_5-q_5; \vec 0 ; -1,1) $ to 
 $(q_1^{\prime}, q_{5}^{\prime} )$ with 
 $ q_1^{\prime} q_5^{\prime} = (Q_1-q_1)(Q_5-q_5) - 1 $, 
 obtaining a $U(q_5^{\prime})$. Generically 
 $q_5^{\prime} + \tilde q_5 \ne Q_5$, so we cannot 
 think of this in terms of something happening 
 within $U(Q_5)$ gauge theory. It is intriguing 
 that whereas all the other splitting processes  
 could be understood, at least roughly,  in terms of some apprpriate
 compactification of instanton moduli spaces, 
 we here have to go beyond gauge theory at a fixed rank. 
 It should be interesting to explore  the  systematic use of 
 gauge theories of arbitrary rank to understand 
 the full physics of the $(Q_1,Q_5)$ system, using duality 
 along the above lines.

\newsec{ Summary and Outlook } 

  We found that counting functions related to 
  the number of splittings of a string in six dimensions
  contains interesting information 
  on symmetries associated with the phenomena 
 of long strings in ADS3 and the associated continuums
 of the spectrum.  The Weyl sub-group 
 of $O(5,5;Z)$ played an interesting role. 

  The nature of the BPS splittings also 
  gives information about the structure 
  of the compactifications of instanton  moduli spaces. 
  The symmetries of BPS splittings are related to
  symmetries acting on strata of the  compactifications.  
  Some puzzles were raised regarding the details 
  of this correspondence. It would be very interesting 
 if these puzzles can be solved and the Fock space structures
 we described can be derived from appropriate 
 compactifications of ( possibly $ \alpha^{\prime}$ and $g_s$-corrected )   
 instanton moduli spaces, thus generalizing the familiar
 relations between Fock spaces and instanton moduli 
 spaces \vafwit. While we focused here on $T^4$, the discussion 
 is easily generalized to $K3$ and similar connections 
 to instanton moduli spaces on $K3$ should exist.

  When we consider splittings of the $(Q_1,Q_5)$ system 
  which include vectors carrying non-trivial NS charges,  
   we need to go beyond conventional compactifications 
  of instanton moduli spaces to even get a rough 
 gauge theoretic understanding of the splittings. 
 One avenue is to use facts about the duality group 
 $O(5,5;Z)$ to show that  we may need to go beyond 
 instanton moduli spaces for bundles of a fixed rank 
 to get a full gauge theory understanding of the 
 $(Q_1,Q_5)$ system.

 Whereas we used the counting of long strings 
 to identify symmetries of the physics of long strings. 
 It will be interesting to explore the consequences
 of these symmetries for the detailed 
 dynamics, e.g the thresholds where the density 
 of continuum states jumps. 
 The counting problems themselves allow 
 refinements mentioned in section 3. 
 It would be interesting to explore if with these refinements, 
 and with generalization to the three-charge 
 system relevant to black hole entropy \stvaf\calma, 
 one can 
 gain new insights into the statistics of black holes.

\noindent{\bf Acknowledgements:}
  We are happy to thank for
   discussions George Daskalopoulos,  Zack Guralnik, 
    Pei Ming Ho, Antal Jevicki,  David Lowe, 
 Juan Maldacena and  Andy Strominger.
 M. Mihailescu was partially supported by a fellowship from 
 the Galkin foundation 
 while this work was being done.  This research was supported by 
 DOE grant  DE-FG02/19ER40688-(Task A).

\listrefs

\end